\documentclass[11pt,sigplan,nonacm]{acmart}
\usepackage[utf8]{inputenc}
\usepackage{ffcode}
\usepackage{natbib}
\usepackage{paralist}
\usepackage{href-ul}
\usepackage{doi}
\usepackage{to-be-determined}
\usepackage{bibcop}

\usepackage{silence}
\title{CAM: A Collection of Snapshots of GitHub Java Repositories Together with Metrics}
\author{Yegor Bugayenko}
\orcid{0000-0001-6370-0678}
\email{yegor256@huawei.com}
\affiliation{\institution{Huawei, Russia, Moscow}\city{}\country{}}

\newcommand\cam{{\sffamily\bfseries CAM}}

\begin{document}

\begin{abstract}
Even though numerous researchers require stable data\-sets along with source code and basic metrics calculated on them, neither GitHub nor any other code hosting platform provides such a resource. Consequently, each researcher must download their own data, compute the necessary metrics, and then publish the dataset somewhere to ensure it remains accessible indefinitely. Our \cam{} (stands for ``Classes and Metrics'') project addresses this need. It is an open-source software capable of cloning Java repositories from GitHub, filtering out unnecessary files, parsing Java classes, and computing metrics such as Cyclomatic Complexity, Halstead Effort and Volume, C\&K metrics, Maintainability Metrics, LCOM5 and HND, as well as some Git-based Metrics. At least once a year, we execute the entire script, a process which requires a minimum of ten days on a very powerful server, to generate a new dataset. Subsequently, we publish it on Amazon S3, thereby ensuring its availability as a reference for researchers. The latest archive of 2.2Gb that we published on the 2nd of March, 2024 includes 532K Java classes with 48 metrics for each class.
\end{abstract}

\maketitle

\section{Motivation}

First, research projects that analyze Java code usually extract it from repositories where open-source projects store their files, such as GitHub. It is common practice in papers explaining results to fully disclose the coordinates of the open-source code being extracted. However, source code is inherently volatile: repositories change their locations and files are modified, as demonstrated by \citet{5463348}. To ensure the replicability of their research results, paper authors must somehow guarantee that the source code used at the time of research remains available and intact throughout the paper's lifetime. One obvious solution would be to make copies of the repositories being extracted and then host them somewhere they are "forever" available.

Second, research methods typically involve filtering out certain types of files found in repositories, such as plain text documents or graphic images, which are not source code. Additionally, some source code files may need to be excluded because they are auto-generated or contain unparseable Java code, making them unsuitable for most methods of code analysis.

Third, most source code analysis research involves collecting metrics from the files found in extracted repositories, such as lines of code, complexity, cohesion, and so on. Most of these metrics are already known, and their retrieval mechanisms are trivial, as summarized by \citet{nunez2017source}.

Thus, there is an obvious duplication of work among different research projects:
\begin{inparaenum}[(a)]
\item they have to ``host'' extracted data to ensure desirable replicability, as noted by \citet{7887704},
\item they must implement filtering of source code fetched from GitHub, and
\item they have to collect popular metrics.
\end{inparaenum}
Having a ready-to-use archive of downloaded, filtered, and measured source code files would help many research projects reduce the amount of work required.

\section{Methodology}

In order to help research projects in all three tasks mentioned above, we created \cam{}\footnote{\url{https://github.com/yegor256/cam}} archive: an open-source collection of scripts regularly (at least once a year) being executed in Docker containers in our proprietary computing environment with results published in form of an ``immutable'' ZIP archive as either a GitHub ``asset'' attached to the next release of our GitHub repository or an object in Amazon S3 (depending on the size of the archive). Here, immutability is not technically guaranteed but promised: even though we, being the owners of the repository, are able to replace any previously created assets, we are not going to do so in order to not jeopardize the idea. Instead, new releases will be published retaining previously generated assets unmodified.

At the time of writing, our GitHub repository consists of scripts written in Makefile, Python, Ruby, and Bash, which do exactly the following:
\begin{itemize}
    \item Fetch 1000 repositories from GitHub, which have \ff{java} language tag, have more than 1K and less than 10K stars, and are at least as big as 200Kb;
    \item Remove files without \ff{.java} extension, Java files with syntax errors, \ff{package-info.java} files, \ff{module-info.java} files, files with lines longer than 1024 characters, and unit tests;
    \item Calculate KLoC, NCSS, Cyclomatic Complexity~\citep{mccabe1976complexity}, Cognitive Complexity~\citep{campbell2018cognitive}, LCOM5~\citep{henderson1996coupling}, NHD~\citep{counsell2006interpretation}, TCC~\citep{bieman1995cohesion}, number of attributes, number of constructors, number of methods, number of static methods, and some other metrics.
\end{itemize}

The size of the latest archive generated on the 2nd of March, 2024 is 2.2Gb. It includes 532K Java classes with 48 metrics for each class. It took us 10 days on a server with eight vCPU and 32Gb of RAM to generate the data.

\section{Limitations}

As of January 2023, \citet{dohmke2023} reported that GitHub hosts more than 420 million repositories, including at least 28 million public repositories, which is the world's largest source code host as of June 2023. According to \citep{daigle2023}, Java is the 4th most popular language on GitHub. Thus, it is reasonable to assume that there are millions of Java repositories on GitHub. It is technically impossible to download and parse even a few percent of this huge data source. In the \cam{} project, we download and scan only a thousand repositories (planning to download a few thousand in the future). Such a tiny fraction of the entire possible scope of analysis is obviously not representative enough. Researchers must understand this limitation and only use \cam{} when representability of the entire Java domain is not the goal of the research.

Even though most of the metrics that we collect have formal definitions given in the papers where the metrics were originally introduced, for example NHD~\citep{counsell2006interpretation} and TCC~\citep{bieman1995cohesion}, there are certain modifications that we had to make to their original algorithms. This happened mostly because modern Java classes have certain features that were not present when said metrics were introduced. Researchers must understand that the metrics generated by the scripts in \cam{} are not exactly the same metrics that were described by their authors.

Even though our scripts download only reasonably popular Java repositories, some of them contain Java files with broken syntax. Also, some files use new Java syntax introduced only in recent versions of Java (such as, for example, ``records'' introduced in Java~21). The parser\footnote{\url{https://github.com/c2nes/javalang}} that we use in \cam{} is only capable of parsing Java~8. We simply exclude all files that are not parseable by this parser. Researchers who are looking for the most current syntax of Java must remember this limitation and try to find another source of data.

\section{Conclusion}

We expect \cam{} archives to be used by research teams analyzing Java source, which want
\begin{inparaenum}[(a)]
\item to guarantee replicability of their results
and
\item to reduce data pre-processing efforts.
\end{inparaenum}
We also expect open-source community to contribute to \cam{} scripts, making filtering more powerful and adding more code metrics to the collection.

{\raggedright
\bibliographystyle{ACM-Reference-Format}
\bibliography{main}}


\begin{thebibliography}{10}


\ifx \showCODEN    \undefined \def \showCODEN     #1{\unskip}     \fi
\ifx \showDOI      \undefined \def \showDOI       #1{#1}\fi
\ifx \showISBNx    \undefined \def \showISBNx     #1{\unskip}     \fi
\ifx \showISBNxiii \undefined \def \showISBNxiii  #1{\unskip}     \fi
\ifx \showISSN     \undefined \def \showISSN      #1{\unskip}     \fi
\ifx \showLCCN     \undefined \def \showLCCN      #1{\unskip}     \fi
\ifx \shownote     \undefined \def \shownote      #1{#1}          \fi
\ifx \showarticletitle \undefined \def \showarticletitle #1{#1}   \fi
\ifx \showURL      \undefined \def \showURL       {\relax}        \fi
\providecommand\bibfield[2]{#2}
\providecommand\bibinfo[2]{#2}
\providecommand\natexlab[1]{#1}
\providecommand\showeprint[2][]{arXiv:#2}

\bibitem[Bieman and Kang(1995)]%
        {bieman1995cohesion}
\bibfield{author}{\bibinfo{person}{James~M. Bieman} {and}
  \bibinfo{person}{Byung-Kyoo Kang}.} \bibinfo{year}{1995}\natexlab{}.
\newblock \showarticletitle{{Cohesion and Reuse in an Object-Oriented System}}.
\newblock \bibinfo{journal}{\emph{{SIGSOFT Software Engineering Notes}}}
  \bibinfo{volume}{20}, \bibinfo{number}{SI} (\bibinfo{year}{1995}),
  \bibinfo{pages}{259--262}.
\newblock
\urldef\tempurl%
\url{https://doi.org/10.1145/223427.211856}
\showDOI{\tempurl}


\bibitem[Campbell(2018)]%
        {campbell2018cognitive}
\bibfield{author}{\bibinfo{person}{G.~Ann Campbell}.}
  \bibinfo{year}{2018}\natexlab{}.
\newblock \showarticletitle{{Cognitive Complexity: An Overview and
  Evaluation}}. In \bibinfo{booktitle}{\emph{{Proceedings of the International
  Conference on Technical Debt}}}. \bibinfo{pages}{57--58}.
\newblock
\urldef\tempurl%
\url{https://doi.org/10.1145/3194164.3194186}
\showDOI{\tempurl}


\bibitem[Cosentino et~al\mbox{.}(2017)]%
        {7887704}
\bibfield{author}{\bibinfo{person}{Valerio Cosentino},
  \bibinfo{person}{Javier~L. Cánovas~Izquierdo}, {and} \bibinfo{person}{Jordi
  Cabot}.} \bibinfo{year}{2017}\natexlab{}.
\newblock \showarticletitle{{A Systematic Mapping Study of Software Development
  With GitHub}}.
\newblock \bibinfo{journal}{\emph{{IEEE Access}}}  \bibinfo{volume}{5}
  (\bibinfo{year}{2017}), \bibinfo{pages}{7173--7192}.
\newblock
\urldef\tempurl%
\url{https://doi.org/10.1109/ACCESS.2017.2682323}
\showDOI{\tempurl}


\bibitem[Counsell et~al\mbox{.}(2006)]%
        {counsell2006interpretation}
\bibfield{author}{\bibinfo{person}{Steve Counsell}, \bibinfo{person}{Stephen
  Swift}, {and} \bibinfo{person}{Jason Crampton}.}
  \bibinfo{year}{2006}\natexlab{}.
\newblock \showarticletitle{{The Interpretation and Utility of Three Cohesion
  Metrics for Object-Oriented Design}}.
\newblock \bibinfo{journal}{\emph{{ACM Transactions on Software Engineering and
  Methodology (TOSEM)}}} \bibinfo{volume}{15}, \bibinfo{number}{2}
  (\bibinfo{year}{2006}), \bibinfo{pages}{123--149}.
\newblock
\urldef\tempurl%
\url{https://doi.org/10.1145/1131421.1131422}
\showDOI{\tempurl}


\bibitem[Daigle(2023)]%
        {daigle2023}
\bibfield{author}{\bibinfo{person}{Kyle Daigle}.}
  \bibinfo{year}{2023}\natexlab{}.
\newblock \bibinfo{title}{{Octoverse: The state of open source and rise of AI
  in 2023}}.
\newblock
  \bibinfo{howpublished}{\url{https://github.blog/2023-11-08-the-state-of-open-source-and-ai/}}.
\newblock
\newblock
\shownote{[Online; accessed 13-03-2024]}.


\bibitem[Dohmke(2023)]%
        {dohmke2023}
\bibfield{author}{\bibinfo{person}{Thomas Dohmke}.}
  \bibinfo{year}{2023}\natexlab{}.
\newblock \bibinfo{title}{{100 million developers and counting}}.
\newblock
  \bibinfo{howpublished}{\url{https://github.blog/2023-01-25-100-million-developers-and-counting/}}.
\newblock
\newblock
\shownote{[Online; accessed 13-03-2024]}.


\bibitem[Henderson-Sellers et~al\mbox{.}(1996)]%
        {henderson1996coupling}
\bibfield{author}{\bibinfo{person}{Brian Henderson-Sellers},
  \bibinfo{person}{Larry~L. Constantine}, {and} \bibinfo{person}{Ian~M.
  Graham}.} \bibinfo{year}{1996}\natexlab{}.
\newblock \showarticletitle{{Coupling and Cohesion (Towards a Valid Metrics
  Suite for Object-Oriented Analysis and Design)}}.
\newblock \bibinfo{journal}{\emph{{Object Oriented Systems}}}
  \bibinfo{volume}{3}, \bibinfo{number}{3} (\bibinfo{year}{1996}),
  \bibinfo{pages}{143--158}.
\newblock


\bibitem[McCabe(1976)]%
        {mccabe1976complexity}
\bibfield{author}{\bibinfo{person}{Thomas~J. McCabe}.}
  \bibinfo{year}{1976}\natexlab{}.
\newblock \showarticletitle{{A Complexity Measure}}.
\newblock \bibinfo{journal}{\emph{{IEEE Transactions on Software Engineering}}}
  \bibinfo{number}{4} (\bibinfo{year}{1976}), \bibinfo{pages}{308--320}.
\newblock
\urldef\tempurl%
\url{https://doi.org/10.1109/TSE.1976.233837}
\showDOI{\tempurl}


\bibitem[Nu{\~n}ez-Varela et~al\mbox{.}(2017)]%
        {nunez2017source}
\bibfield{author}{\bibinfo{person}{Alberto~S. Nu{\~n}ez-Varela},
  \bibinfo{person}{H{\'e}ctor~G. P{\'e}rez-Gonzalez},
  \bibinfo{person}{Francisco~E. Mart{\'\i}nez-Perez}, {and}
  \bibinfo{person}{Carlos Soubervielle-Montalvo}.}
  \bibinfo{year}{2017}\natexlab{}.
\newblock \showarticletitle{{Source Code Metrics: A Systematic Mapping Study}}.
\newblock \bibinfo{journal}{\emph{{Journal of Systems and Software}}}
  \bibinfo{volume}{128} (\bibinfo{year}{2017}), \bibinfo{pages}{164--197}.
\newblock
\urldef\tempurl%
\url{https://doi.org/10.1016/j.jss.2017.03.044}
\showDOI{\tempurl}


\bibitem[Robles(2010)]%
        {5463348}
\bibfield{author}{\bibinfo{person}{Gregorio Robles}.}
  \bibinfo{year}{2010}\natexlab{}.
\newblock \showarticletitle{{Replicating MSR: A Study of the Potential
  Replicability of Papers Published in the Mining Software Repositories
  Proceedings}}. In \bibinfo{booktitle}{\emph{{IEEE Working Conference on
  Mining Software Repositories}}}. \bibinfo{pages}{171--180}.
\newblock
\urldef\tempurl%
\url{https://doi.org/10.1109/MSR.2010.5463348}
\showDOI{\tempurl}


\end{thebibliography}

\end{document}